\documentclass[aps,prl,twocolumn,superscriptaddress,amsmath,floatfix]{revtex4-2}

\usepackage{graphicx}
\usepackage{dcolumn}
\usepackage{bm}
\usepackage{amssymb}
\usepackage{hyperref}
\usepackage{multirow}
\usepackage{color}
\usepackage{enumitem}

\begin{document}

\title{The Kovacs memory effect in a thin granular layer: experimental evidence and
  its physical origin}

\author{Francisco Vega Reyes}
\email{fvega@eaphysics.xyz}
\affiliation{Departamento de F\'{\i}sica and Instituto de
  Computaci\'on Cient\'fica Avanzada (ICCAEx), Universidad de
  Extremadura, 06071 Badajoz, Spain}
\author{\'Alvaro Rodr\'iguez-Rivas}
\email{arodriguezrivas@us.es}
\affiliation{Departamento de Matem\'atica Aplicada II, Universidad de Sevilla, E.T.S. de Ingenier\'{\i}a, C. de los Descubrimientos, s/n. Pabell\'on Pza. de Am\'erica 41092 Sevilla (Spain)}
\author{Pablo Maynar}
\email{maynar@us.es} 
\affiliation{F\'isica Te\'orica, Universidad de Sevilla, Apartado de Correos 1065, 41080 Sevilla, Spain}
\author{M. Isabel Garc\'ia de Soria}
\email{gsoria@us.es}
\affiliation{F\'isica Te\'orica, Universidad de Sevilla, Apartado de Correos 1065, 41080 Sevilla, Spain}

\date{\today}

\begin{abstract}
  We report the experimental observation of memory effects in a vertically vibrated thin granular layer. Following
  a quench in the input acceleration, the granular temperature exhibits an anomalous Kovacs memory effect
  confined to the initial fast relaxation stage. This memory vanishes shortly thereafter, yielding a
  time-dependent memoryless regime governed solely by the instantaneous temperature before the system reaches
  its final steady state. We develop a kinetic theory framework that quantitatively captures these features by
  identifying the initial memory and subsequent memoryless regimes with the kinetic and hydrodynamic states,
  respectively (that are well established in kinetic theory). Our analysis reveals that memory emerges during
  fast transients through coupling between horizontal and vertical temperatures—a mechanism that fundamentally
  constrains the accessible memory phenomenology and precludes observation of the standard Kovacs effect in
  this system. Molecular dynamics simulations provide independent confirmation of all experimental and
  theoretical findings.
\end{abstract} 

\maketitle

\emph{Introduction}---In physical systems, \emph{memory} refers to the ability to encode information about
past states in a manner that influences subsequent evolution \cite{Keim_2019}. The study of memory formation
has attracted growing interest due to its broad relevance across nonequilibrium physics. Within this context,
two phenomena have been studied extensively: the Mpemba and Kovacs effects. The Mpemba effect describes
anomalous relaxation whereby a hotter (or colder) system may reach thermal equilibrium faster than a cooler
(or hotter) one \cite{Lasanta_2017,Lu_2017}. The Kovacs effect, initially observed in polymers
\cite{Kovacs_1979} and later extended to active and granular matter
\cite{Kursten_2017,Prados_2014,Trizac_2014,bgmb2014,Santos_2025}, levitated nanoparticles \cite{Militaru_2021} and other systems, manifests as a system's tendency to transiently
revert toward a past state during relaxation toward a stationary state.

\begin{figure}[ht]
  \includegraphics[width=0.95\columnwidth,clip]{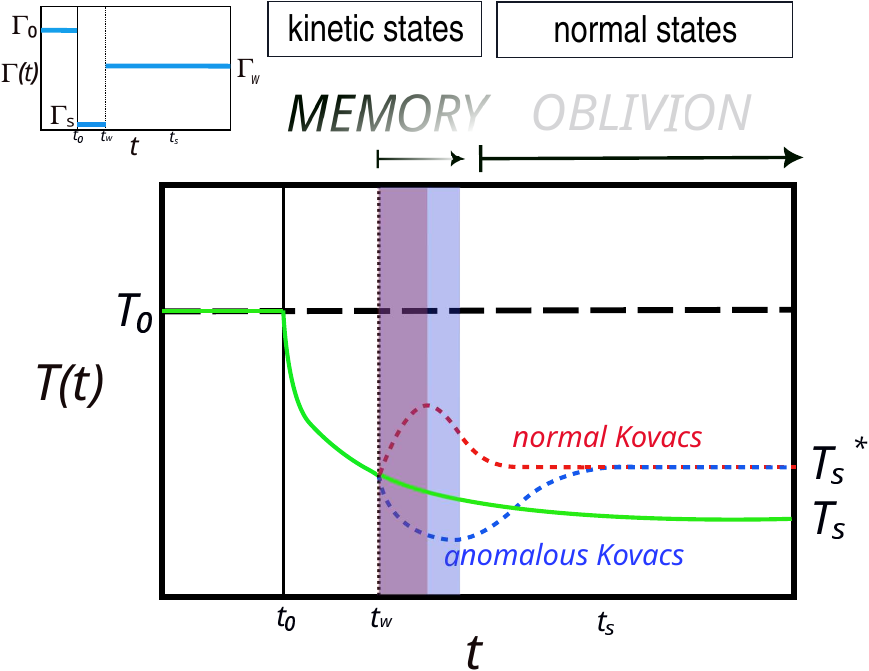}
  \caption{Schematic representation of the Kovacs cooling protocol and the two kinds of memory effects
    considered. The granular fluid initially at stationary temperature $T_0$ (the energy input being
    $\Gamma_0$) is quenched by lowering the energy input to $\Gamma_s$ at time $t_0$ and relaxes monotonically
    toward a new stationary state $T_s < T_0$, that would be reached at a long time $t_s$ (green solid
    line). However, at an intermediate waiting time $t_w$ (where $t_0 < t_w< t_s$), the energy injection is
    adjusted to $\Gamma_w$ so that the new stationary temperature $T_s^*$ equals the instantaneous
    temperature: $T(t_w) = T_s^* > T_s$. The normal Kovacs effect is produced in the form of an upwards hump
    of the temperature, and hence
    $\mathrm{sign}(\mathrm{d}T/\mathrm{d}t|_{t_w^-}) \neq \mathrm{sign}(\mathrm{d}T/\mathrm{d}t|_{t_w^+})$. In
    this case, the system transitorily tends to reach the initial state that the system had at $t=t_0$ (red
    dot line). On the contrary, the anomalous Kovacs effect is characterized by a downwards hump and now
    $\mathrm{sign}(\mathrm{d}T/\mathrm{d}t|_{t_w^-}) = \mathrm{sign}(\mathrm{d}T/\mathrm{d}t|_{t_w^+})$ where
    the system tries to continue cooling down (blue dot line), as if it remembered the previous relaxation
    process instead. In both cases, the system remembers features of the past.  The top left inset shows the
    corresponding changes of the input signal $\Gamma$.  The blue and red shaded regions (red appears as
    purple due to overlap with blue) signal the time intervals where the memory effect takes
    place. \label{fig:Kovacs_types} }
\end{figure}

Granular fluids provide a compelling platform for exploring such memory effects. Previous experimental work
has examined memory in the compaction of shaken granular layers \cite{Josserand_2000} and the Mpemba and
Kovacs effects have been theoretically studied in driven granular gases \cite{Lasanta_2017,Lasanta_2019}. The
Kovacs effect appears as non-monotonic evolution of an observable---typically volume, stress, or
temperature---upon resuming relaxation after an intermediate perturbation. Notably, far-from-equilibrium
analyses predict an \emph{anomalous} Kovacs response in which relaxation initially proceeds in the same
direction as before the perturbation, driving the system further from its instantaneous target state
\cite{Prados_2014,Lasanta_2019}; see Fig.~\ref{fig:Kovacs_types}.

Despite these theoretical developments, experimental evidence of memory effects in granular fluids remains
scarce, though the Mpemba effect has been reported in other systems
\cite{Gianluca_2025}. In this Letter, we present the first experimental observation of a Kovacs effect in
  a granular fluid, following the standard protocol. In particular, we reveal a
robust memory signature associated with the granular temperature
(i.e., the average particle kinetic energy) \cite{Brey_1998}. In
contrast to earlier studies 
focused on density \cite{Josserand_2000} or stress
\cite{Candela_2023}, our results demonstrate memory effects 
directly through thermal relaxation dynamics.


Since particle collisions are inelastic, a steady state in granular matter can only be achieved through
sustained energy input, the intensity of which is here described by the parameter $\Gamma$ \cite{wm96,
  plmpv98, clh2000, brs2013}. The thermal Kovacs cooling protocol proceeds as follows \cite{Kovacs_1979,
  Prados_2014, bgmb2014}: a granular fluid initially at stationary temperature $T_0$ (the energy input being
$\Gamma_0$) is quenched (by lowering the energy input to $\Gamma_s$) at time $t_0$ and relaxes monotonically
toward a new stationary state $T_s < T_0$, that would be reached at a long time $t_s$. However, at an
intermediate waiting time $t_w$ (where $t_0 < t_w< t_s$), the energy injection is adjusted to
$\Gamma_w > \Gamma_s$ so that the new stationary temperature $T_s^*$ equals the instantaneous temperature:
$T(t_w) = T_s^* > T_s$ \cite{Mompo_2021,Lasanta_2019}. In a \emph{normal} Kovacs response,
$\mathrm{d}T/\mathrm{d}t$ reverses sign at $t_w$, producing an upward hump. In an \emph{anomalous} response,
the sign is preserved and the system continues cooling, generating a downward hump
(Fig.~\ref{fig:Kovacs_types}) \cite{Prados_2014, bgmb2014}.
 
The Kovacs effect may appear paradoxical since it can occur even when all hydrodynamic variables equal their
stationary values \cite{Kovacs_1979}. Memory is reflected in a non-monotonic behavior of the temperature,
implying that, at those times, the temperature does not fully specify the state of the system. Kinetic theory
explains this behavior through the distinction between \emph{kinetic} and \emph{hydrodynamic} regimes
\cite{Hilbert,C69,C70,B72,Brey_1998}: in the latter, the memory kernel arising from the elimination of
higher-order moments has collapsed to an effectively local form, and the reduced description becomes
Markovian. Moreover, the dynamics of the velocity distribution function consists of a rapid
and initial-condition-dependent evolution (the kinetic regime) over a characteristic time $t_m$, followed by a
hydrodynamic regime in which all the time dependence enters solely through the hydrodynamic fields, the
granular temperature in our case.  During the kinetic stage ($t\lesssim t_m$), the velocity distribution
function depends on both $T$ and higher-order moments $\{a,b,\dots\}$:
\begin{equation}
  \label{eq:f_kinetic}
f(\mathbf{v},t) = f(\mathbf{v}\mid T(t), a(t), b(t), \dots), 
\end{equation}
while in the hydrodynamic regime ($t\gtrsim t_m$), the distribution simplifies to
\begin{equation} 
  \label{eq:f_normal}
f(\mathbf{v}, t) = f(\mathbf{v}\mid T(t)),
\end{equation} 
where the system cannot return to kinetic states unless perturbed anew \cite{B72,Hilbert}. Clearly, the
hydrodynamic fields specify the state of the system only in the hydrodynamic regime. In this regime, the
temperature evolves autonomously and all other moments are slaved to it, so that the macroscopic dynamics
becomes independent of the prior kinetic history (memory vanishes). In the kinetic regime, even when $T=T_s$,
evolution of the moments $\{a,b,\dots\}$ can drive further changes in $T$. In the Kovacs protocol described
above, the system is typically in the hydrodynamic regime just before the quench and it is automatically
driven out of it after the quench and memory effects may arise.

In this letter, we consider a thin vertically vibrated granular layer in which the walls move sinusoidally
with amplitude, $A$, and angular frequency, $\omega$.  The energy input is thus measured by the mean
dimensionless acceleration of the walls, $\Gamma\equiv A\omega^2/g$, with $g$ being the gravitational
acceleration. The base state of the fluid phase in this kind of system is homogeneous and the dynamics is
quasi-2D \cite{Melby_2005}. Thus, the relevant hydrodynamic field is the horizontal granular temperature,
$T = (m/2) \langle v_x^2 + v_y^2\rangle$ \cite{Olafsen_1998,Melby_2005} (where $\langle\dots\rangle$ denotes
particle average, $m$ is particle mass and $\mathbf{v}$ is particle velocity). By combining theory,
simulations, and experiments, we demonstrate that (i) the Kovacs effect in vibrated thin granular layers is
intrinsically anomalous, (ii) memory is confined to kinetic stages, and (iii) subsequent relaxation proceeds
through a universal memoryless hydrodynamic state governed solely by the horizontal temperature. In addition,
the results show that memory characteristics are constrained by the system's governing equations and cannot be
tuned through external conditions alone.

\begin{figure}[ht]
  \includegraphics[width=0.99\columnwidth,clip]{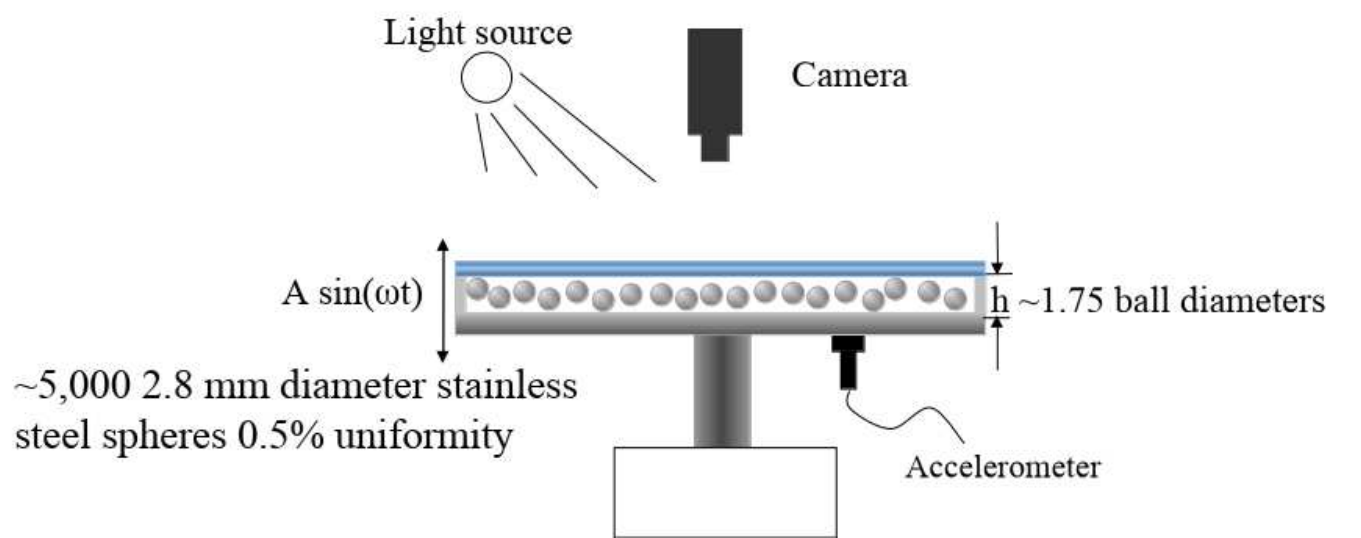}
  \caption{Schematic representation of the experimental setup. The thin granular layer of steel spheres is
    vertically vibrated. The input acceleration $\Gamma = A\omega^2/g$ is  monitored with an accelerometer,
    where $A$ is the vibration amplitude, $\omega$ the angular frequency, and  $g=9.8~\mathrm{m/s^2}$. 
    \label{fig:exp_set-up_sketch}} 
\end{figure}

\emph{Experimental configuration}- Our experimental setup follows the design used in previous works by Urbach
and co-workers \cite{Olafsen_1999,Melby_2005}; see Fig.~\ref{fig:exp_set-up_sketch}. It consists of a thin,
circular layer of $N$ identical stainless-steel spheres (AISI~52100), with coefficient of normal restitution
$\alpha=0.95$ \cite{Louge_1999}. The spheres diameter is $\sigma \simeq 2.38~\mathrm{mm}$. The layer is
confined between a bottom aluminum plate and a top plexiglas lid, with steel spacers forming the lateral
boundary. The separation between the plates, $h$, satisfies $\sigma < h < 2 \sigma$, ensuring that particles
cannot overlap while still allowing collisions with both vertical and horizontal components of the relative
velocity. This geometry promotes efficient energy redistribution when the system is vertically vibrated by an
electromagnetic shaker (the input signal has a form $A\sin(\omega t)$) and allows for horizontal particle
position and velocity measurement \cite{Crocker_1996,Olafsen_1998}.

All experiments were performed at a driving frequency $\nu \equiv\frac{\omega}{2\pi}=180~\mathrm{Hz}$. The
layer height was set to $h = 1.75\,\sigma$. The two-dimensional density was fixed at $\rho_{2D}=0.45$, which
corresponds to $N = 4711$ particles. Here, $\rho_{2D}$ is defined following $\rho_{2D}\equiv N/N_\mathrm{max}$
~\cite{Melby_2005,Reyes_2008} with $N_{\mathrm{max}} = 10469$ being the maximum number of particles fitting
into the circular domain under close packing.  For additional details, including a list of the equipment used,
see the Appendix A.

\emph{Molecular dynamics simulations}- We performed molecular dynamics simulations, following an algorithm
analogous to the one we used in previous works \cite{Melby_2005,Reyes_2008}. Wall-particle and
particle-particle collisions are inelastic and we model them by using a viscoelastic collisional model
\cite{Melby_2005}. The walls move sinusoidally as in the experiment. Frequency and vibration amplitude may be
controlled independently, as well as other relevant parameters in the system. For more details, see Appendix
B.

\begin{figure}[ht]   
   \includegraphics[width=0.9\columnwidth,clip]{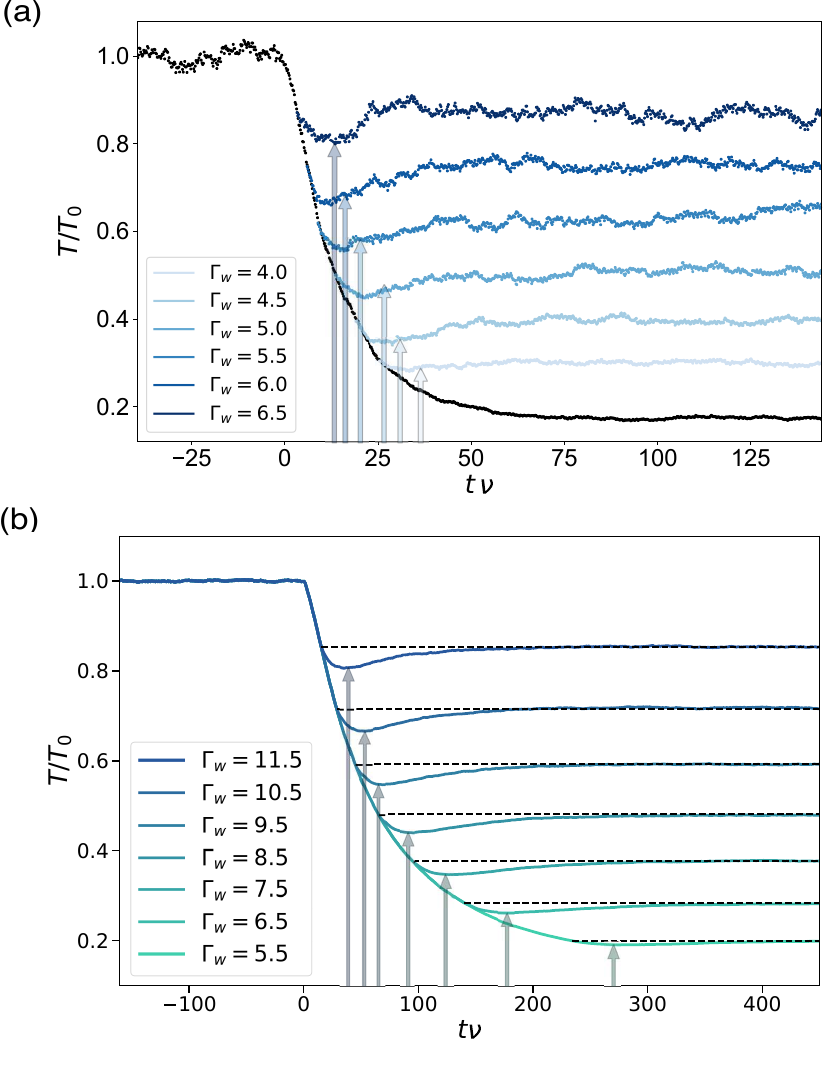}
   \caption{ Horizontal temperature (blue) relaxation curves for a Kovacs protocol in experiments and MD
     simulations, in reduced magnitudes. (a) Experiments: initial and final acceleration inputs are
     $\Gamma_0 = 7.0$ and $\Gamma_s = 3.15$ respectively, while intermediate values of the input acceleration
     at different waiting times $t_w$ are $\Gamma_w = \{6.5; 6.0; 5.5; 5.0; 4.5; 4.0\}$.
     $T_0 = T_s(\Gamma = 7.0)$; $\nu = 180~\mathrm{Hz}$ (b) Molecular dynamics: $\nu =
     643.217~\mathrm{Hz}$. $\Gamma_0 = 12.5$ amd $\Gamma_s = 5.0$ respectively, and
     $\Gamma_w = \{11.5; 10.5; 9.5; 8.5; 7.5; 6.5; 5.5\}$. For both experiment and simulation, darker blue in
     the curves corresponds to higher $\Gamma_w$. Arrows mark the reduced $t_m$ values extracted from the time
     evolution of the kurtosis. $T_0 = T_s(\Gamma = 12.5)$. \label{fig:kovacs_results} }
 \end{figure}

\emph{Results}--- Through extensive experimental measurements and molecular dynamics (MD) simulations, we
verify that our system exhibits exclusively the \emph{anomalous} Kovacs effect
(Figure~\ref{fig:kovacs_results}). Following the protocol described above, after each quench at different
waiting times $t_w$, the system will ultimately evolve towards stationary temperatures, $T_s^{*}$, that depend
on $\Gamma_w$. However, prior to this, we observed in all cases that at $t_w$ an extended cooling takes place
during a finite time interval until reaching a minimum temperature $T_\mathrm{min}$ at
$t_\mathrm{min}$. Beyond this point, the temperature increases monotonically, ultimately reaching the steady
value, $T_s^*$, as observed in all relaxation curves in Figure~\ref{fig:kovacs_results}.  To quantify the
memory response amplitude, we define the anomalous Kovacs measure as
$\mathcal{H}_\mathrm{min} = T_\mathrm{min}/T_s^* - 1$. Thus, $\mathcal{H}_\mathrm{min}< 0$ in our system. In
Figure~\ref{fig:kovacs_results}(a) for experiments and Figure~\ref{fig:kovacs_results}(b) for simulations we
can see that the memory amplitude $\mathcal{H}_\mathrm{min}$ tends to decrease as $t_w \to t_s$, and it is
expected to become null for $t_w = t_s$ since in that case the kinetic states have disappeared by definition
(because the system is already in steady state).  A more quantitative analysis of the behavior of the memory
amplitude is performed in the \emph{Theory} section.

We identify the onset of hydrodynamics when the next higher moment (kurtosis 
$K=\langle (v_x^2 + v_y^2)^2\rangle/\langle (v_x^2 +v_y^2) \rangle^2$) becomes stationary.  At this point it
is expected that all remaining time-dependence in the distribution function occurs through the lower-order
moments—specifically, through the hydrodynamic field $T$. The system state thus becomes effectively
hydrodynamic. To verify this picture, we measured the kurtosis $K$ relaxation (see Figure~1 in the
Supplementary Material file for more detail) across all experimental and simulation series. We indicate, by
means of vertical arrows in Figure~\ref{fig:kovacs_results}, the times at which $K$ becomes stationary
(indicated by the arrows). Remarkably, in all cases, the temperature minimum $T_\mathrm{min}$ occurs before
this hydrodynamic transition point. This temporal ordering directly demonstrates that the Kovacs memory
effect—signaled by the non-monotonic temperature evolution—vanishes precisely when hydrodynamic behavior is
established. Therefore, anomalous memory effects cannot persist in truly hydrodynamic states.
 

\begin{figure}[ht]
  \includegraphics[width=0.875\columnwidth,clip]{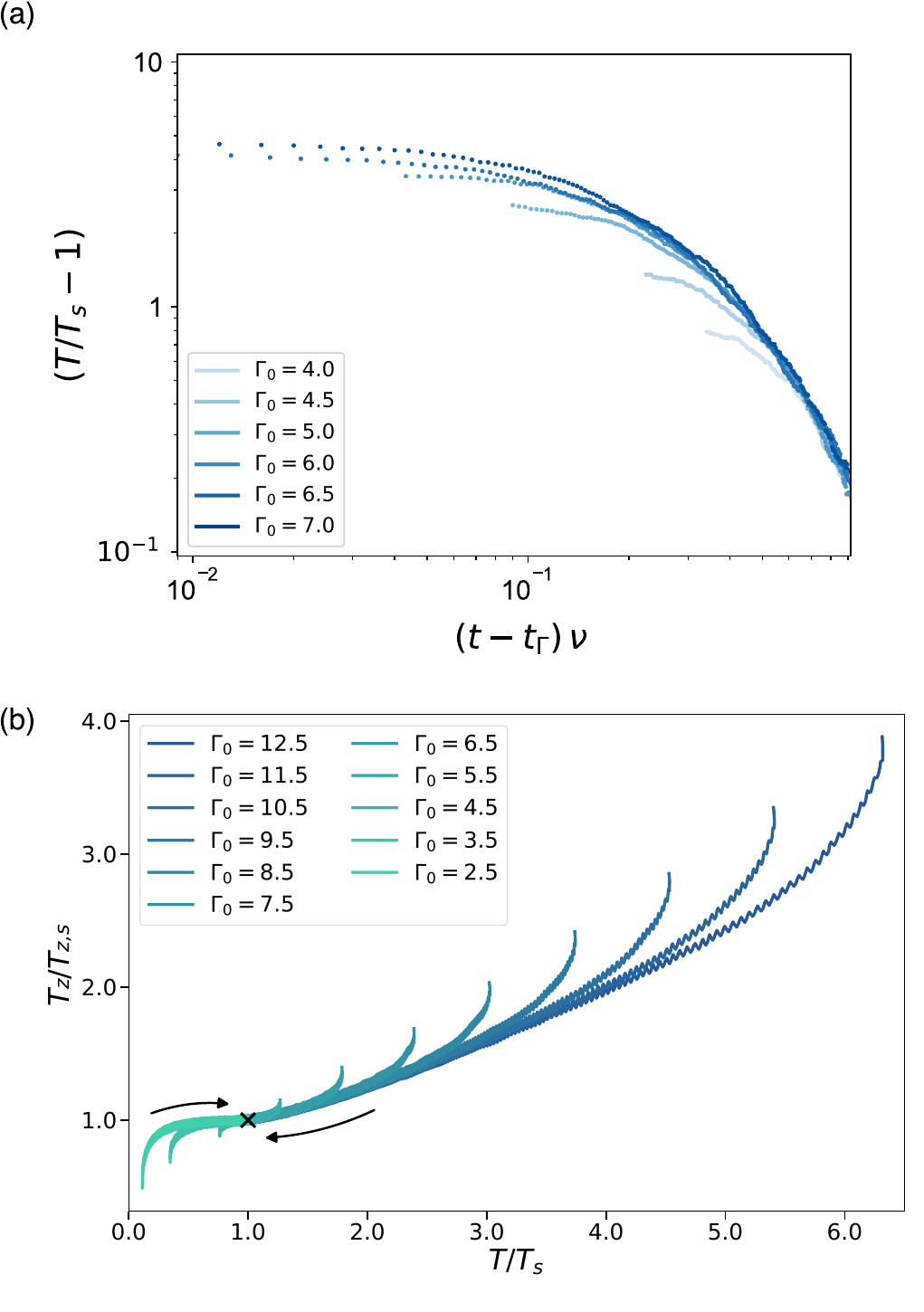}
   \caption{(a) Relaxation curves $T/T_s-1$ vs. $(t-t_{\Gamma})\,\nu$ from experiments, for a step-down protocol with $\Gamma_s =
     3.15$, and $\Gamma_0 = \{6.5; 6.0; 5.5; 5.0; 4.5; 4.0 \}$. Here $t_{\Gamma}$ denotes the translation in
     the
     time performed to each curve to make the curves collapse. $T_s$ stands for the steady horizontal
     temperature for $\Gamma_s = 3.15$. $\nu = 180~\mathrm{Hz}$. (b) Relaxation curves of $T_z/T_{z,s}$ vs. $T/T_s$ from
     simulation data, for a series of values of initial input accelerations
     $\Gamma_0 = \{2.5; 3.5; 4.5; 5.5; 6.5; 7.5; 8.5; 9.5; 10.5; 11.5; 12.5 \}$. $T_s$ and $T_{z,s}$ denote the steady-state horizontal and vertical temperatures, respectively, at the final state $\Gamma_s = 5.0$.  
     \label{fig:Th_relaxation}}
 \end{figure}

 The transition to memoryless hydrodynamic behavior can also be demonstrated directly from the relaxation
 dynamics of the horizontal temperature. We analyze this in Figure~\ref{fig:Th_relaxation}, where the system
 initially at steady states with different $\Gamma_0$, $T_0$, converges to the same stationary temperature
 (due to a quench to the same value $\Gamma_s$). Figure~\ref{fig:Th_relaxation}(a) shows experimentally that,
 after a convenient translation in the time axes, $t_{\Gamma}$, the horizontal temperature curves collapse
 onto a single curve. The temporal translation $t_\Gamma$ arises naturally as the time required for the system
 to leave the initial kinetic transient and reach the hydrodynamic regime, allowing comparison of states with
 the same temperature independently of their preparation. This data collapse demonstrates that $T$ is the
 unique hydrodynamic variable controlling long-time dynamics. The simulations provide additional quantitative
 insight through access to the vertical temperature $T_z= m\langle v_z^2\rangle$.
 Figure~\ref{fig:Th_relaxation}(b) plots $T_z$ versus $T$ for different initial conditions under both heating
 and cooling protocols. After an initial transient, $T_z$ establishes a universal relationship with $T$,
 following a single curve toward a unique stationary state (marked with a cross). This collapse onto a
 universal manifold elegantly demonstrates how non-hydrodynamic modes relax onto the lower-dimensional
 manifold defined by the hydrodynamic variable $T$. Although Figs. 4(a) and 4(b) use different
 representations, both reflect the same underlying mechanism: after a transient, the dynamics collapses onto a
 hydrodynamic manifold parametrized by $T$.


 \emph{Theory}- To explain the above phenomenology, we consider the simple microscopic model introduced in
 \cite{Maynar_2019}: an ensemble of $N$ hard spheres of diameter $\sigma$ that collide inelastically with a
 velocity independent coefficient of normal restitution, $\alpha$, confined between two walls of area $S$
 separated a distance $h<2\sigma$. The top wall is elastic, while the bottom wall mimics a vibrating sawtooth
 wall of mean velocity $v_0$ that injects energy into the system (details are given in the Supplementary
 Material file).  The simplified model captures the relevant physics because it retains the key mechanisms
 present in the experiments: anisotropic energy injection, collisional dissipation, and coupling between
 vertical and horizontal degrees of freedom.  For this model, a kinetic equation can be formulated, i.e., a
 closed equation for the one-particle distribution function, $f(\mathbf{r}, \mathbf{v}, t)$, that accurately
 describes the dynamics of the system. For spatially homogeneous states, the horizontal and vertical
 temperatures are defined as $n T = \int\mathrm{d}\mathbf{v} \frac{m}{2} (v_x^2 + v_y^2)f(\mathbf{v},t),$ and
 $\frac{n}{2} T_z = \int\mathrm{d}\mathbf{v} \frac{m}{2} v_z^2 f(\mathbf{v},t)$, respectively,
where $n=\frac{N}{S(h-\sigma)}$ is the particle density. From the
kinetic equation, assuming that the 
distribution function is a gaussian with different horizontal and vertical
temperatures (in agreement with the results of the previous section), 
closed evolution for the temperatures are obtained to leading order in
$\epsilon=(h-\sigma)/\sigma$, obtaining 

\begin{align}
  \frac{\mathrm{d}T}{\mathrm{d}{t}} &= \sqrt{\frac{\pi T}{m}} (1+\alpha) \varepsilon n
                \sigma^2  \label{eq:dif_Th} \\
                &  \times\left[-(1-\alpha) T + \varepsilon^2 \left(-\frac{5\alpha-1}{12}T + \frac{3\alpha + 1}{12}
                             T_z\right)\right], \nonumber\\
  \frac{\mathrm{d}T_z}{\mathrm{d}t} & = \frac{2}{3} \sqrt{\frac{\pi T}{m}} (1+\alpha) \varepsilon^3 n
                                      \sigma^2 
       \left(\frac{1 + \alpha}{2} T- T_z \right) + \frac{2
         v_0}{\varepsilon \sigma} T_z. \label{eq:dif_T} 
\end{align} 
The structure of the equations is clear: energy is injected in the vertical direction (last term of
Eq. (\ref{eq:dif_T})), while it is transferred to the other degrees of freedom via inelastic collisions. In
the long time limit, a stationary state is reached in which the energy injected by the bottom wall is
compensated by the energy lost in collisions. As the energy is injected in the vertical direction, if the
velocity of the bottom wall is suddenly changed at $t=t_w$,
$\mathrm{d}T/\mathrm{d}t|_{t_w^-} = \mathrm{d}T/\mathrm{d}t|_{t_w^+}$, because $v_0$ does not explicitly
appears in Eq. (\ref{eq:dif_Th}). Hence, for this model, only the anomalous Kovacs effect appears. Moreover,
an analysis of Eqs. (\ref{eq:dif_Th}) and (\ref{eq:dif_T}) indicates that, after a transient, the system
forgets the initial condition and reaches a time-dependent solution in which the vertical temperature is
slaved to the horizontal temperature \cite{Maynar_2019}.  This behavior is consistent with the MD simulations
shown in Fig. 4(b), where the vertical temperature is not constant but becomes a function of the horizontal
one, $T_z = T_z(T)$, in contrast with quantities such as the kurtosis that reach stationary values in the
hydrodynamic regime. The stationary state is reached through this time-dependent universal solution indicating
that the horizontal temperature is the relevant hydrodynamic variable in agreement to what happens in the
experiments (see Fig. \ref{fig:Th_relaxation}(a)). A similar behavior appears in other heated granular models
\cite{GSoria2012}.

Close to the stationary state a more quantitaive analysis can be done by linearizing Eqs. (\ref{eq:dif_Th})
and (\ref{eq:dif_T}) around the stationary state (see SM). To study the shape of the hump, we introduce the
function $\mathcal{H}(s)=T(s)/T_s-1$, where $s$ is proportional to the collisions per particle in $(t_w, t)$,
$\mathrm{d}s = (1+\alpha)\, n \sigma^2 \epsilon\, \sqrt{\frac{\pi T_s}{m}}\, \mathrm{d}t$. In the linear
theory, the function $\mathcal{H}(s)$ can be explicitly calculated, obtaining
\begin{equation}
  \label{eq:theta_vs_t}
  \mathcal{H}(s) = \frac{e^{\lambda_1 s} - e^{\lambda_2 s}}{\lambda_1 - \lambda_2}\,
  \frac{\mathrm{d}\mathcal{H}}{\mathrm{d}s}\bigg|_{s=0}.
\end{equation}
$\lambda_1$ and $\lambda_2$ are two dimensionless negative functions of $\alpha$ and $\epsilon$ (their
explicit expressions are given in the SM) associated to two time scales: a fast scale associated to
$\lambda_2$ and a slow scale (hydrodynamic scale) associated to $\lambda_1$ (actually,
$\lim_{\alpha\to 1}\lambda_1=0$). As $\lambda_2 < \lambda_1<0$,
$\mathrm{sign}\big[\mathcal{H}(s)\big] =
\mathrm{sign}\left[\frac{\mathrm{d}\mathcal{H}}{\mathrm{d}s}|_{s=0}\right] $. Since the cooling Kovacs protocol
requires $\frac{\mathrm{d}\mathcal{H}}{\mathrm{d}s}|_{s=0} < 0$, it follows that $\mathcal{H}(s) < 0$ (see
Figure~\ref{fig:kovacs_results}). Moreover, from Eq. (\ref{eq:theta_vs_t}), it follows that the time interval
to reach the minimum (in the $s$-scale) measured from $t_w$ does not depend on $t_w$ and that the amplitude of
the hump, $\mathcal{H}_\mathrm{min}$, is proportional to
$\frac{\mathrm{d}\mathcal{H}}{\mathrm{d}s}|_{s=0}$. This is in agreement with the results of
Fig. \ref{fig:kovacs_results}, where it is shown that the amplitude of the hump decreases with the input
acceleration.


\emph{Conclusions}---We have provided the first experimental observation of thermal memory in a granular
fluid, manifested as a Kovacs-type effect. In contrast to earlier granular memory observed through density or
stress \cite{Josserand_2000}, here memory is carried by the granular temperature itself. In the confined
granular layer, only the anomalous Kovacs response is observed: during relaxation, the system initially
departs further from its target state, with complete absence of the classical (normal) Kovacs hump. Through a
minimal kinetic theory model and molecular dynamics simulations, we establish that the dynamics is governed by
two coupled temperatures---horizontal and vertical---operating on slow and fast timescales, respectively. This
directional nature of energy injection directly constrains the accessible memory phenomenology.

Our experiments reveal that memory is confined to the initial rapid transient (the kinetic stage). This is a
fundamentally different regime from persistent path-dependent memory in externally-driven systems (such as
magnetic hysteresis). Here, by contrast, memory arises only during the fast transient where both temperature
components are dynamically active, while subsequent evolution follows a universal hydrodynamic state, governed
solely by the horizontal temperature, in which the kurtosis is constant and the temperature curves collapse
(Fig.~\ref{fig:Th_relaxation}).

This confinement of memory to the kinetic stage has a direct interpretation in terms of the Zwanzig-Mori
formalism, in which eliminating degrees of freedom generates memory kernels that become effectively local once
the retained and projected-out variables decouple~\cite{Zwanzig_1960,Mori_1965,Forster_1975}. In effect, while
the kurtosis is still evolving, a description in terms of the granular temperature alone is an incomplete
projection from the full kinetic dynamics, and the memory kernel remains non-local; the Kovacs hump is its
dynamical signature. Stationary kurtosis thus marks the closure of the projection, after which the temperature
evolves autonomously.

Our results indicate that the controllability of memory in driven systems is itself
regime-dependent---determined by whether the projected-out degrees of freedom remain externally accessible:
external engineering of relaxation, as achieved in driven quantum systems \cite{Beato_2026}, requires such
accessibility, whereas in driven granular fluids the memory phenomenology is intrinsically fixed by the
kinetic equations. In the latter case, the kurtosis criterion provides a generic, experimentally accessible
signature for the onset of memoryless hydrodynamic behavior — readily measurable in other driven systems such
as active matter or dense colloidal suspensions.

\begin{acknowledgments}
  We thank Prof. A. Santos for useful comments on the early versions of this work and Profs. J. J. Brey and
  J. S. Urbach for later insightful discussion. We acknowledge support from Ministerio de Ciencia,
  Innovaci\'on y Universidades through Agencia Estatal de Investigación (AEI) through projects
  no. PID2020-116567GB-C22 (FVR), no. PID2024-156257NB-C21 and No. PID2021-126348NB-I00 (MIGS and
  PM) funded by MCIN/AEI/10.13039/501100011033 and ERDF “A way of making Europe''. F.V.R. also acknowledges
  support from Junta de Extremadura through contract No. GR24077 and MIGS and PM from the Grant
  No. ProyExcel-00505, funded by the Junta de Andaluc\'ia. The authors gratefully acknowledge the Centro
  Inform\'atico Cient\'ifico de Andaluc\'ia (CICA), part of the Agencia Digital de Andaluc\'ia (ADA), for
  providing high-performance computing (HPC) resources from the Hércules Cluster and technical support.

\end{acknowledgments} 

\bibliography{kovacs_thin_layer}

\newpage
\onecolumngrid
\vspace{\columnsep}
\begin{center}
\rule[3pt]{0.4\textwidth}{0.4pt}
\textbf{\large{ \ End Matter \ }}
\rule[3pt]{0.4\textwidth}{0.4pt}
\end{center}
\vspace{\columnsep}
\twocolumngrid

\setcounter{equation}{0}
\setcounter{figure}{0}
\setcounter{table}{0}
\renewcommand{\theequation}{A\arabic{equation}}
\renewcommand{\thefigure}{A\arabic{figure}}

\appendix

\section{Appendix A: Experimental Methods}
\label{app:experiment}

The experiments were performed using a Bruel \& Kj\ae r vibration system composed of an amplifier
(LDS~LPA600), controller (Comet~COM200), and electromagnetic shaker (LDS~V406) with an internal amplifier
(PA500L). The system delivers up to $196~\mathrm{N}$ of force and a maximum peak-to-peak displacement of
$17.6~\mathrm{mm}$. A triaxial accelerometer (model~4535-B, sensitivity $10~\mathrm{mV}/g$, range $700~g$)
provides precise closed-loop control of the vertical acceleration. It measures also the horizontal
acceleration, which helps to achieve very accurate horizontality of the thin layer. We checked that input
acceleration fluctuations remain below $1\%$. System operation and monitoring are performed using Br\"uel \&
Kj\ae r software (SCO-02V). The shaker is cooled by a $50~\mathrm{Hz}$ fan to ensure stable performance, and a
low-frequency absorption plug is mounted beneath the plate to suppress low-frequency noise in the
accelerometer input signal.

A Phantom VEO high-speed camera records the particle motion at 1000\,fps. Camera triggering and acceleration
changes are synchronized using a Keysight DSOX2024A oscilloscope (4 channels, 200~MHz bandwidth,
$\sim1~\mathrm{GS/s}$ sampling rate). This arrangement allows frame-level precision in identifying the instant
of acceleration change, which is essential for protocol timing.

\begin{figure}[ht]
  \centering
  \includegraphics[width=0.75\columnwidth,clip]{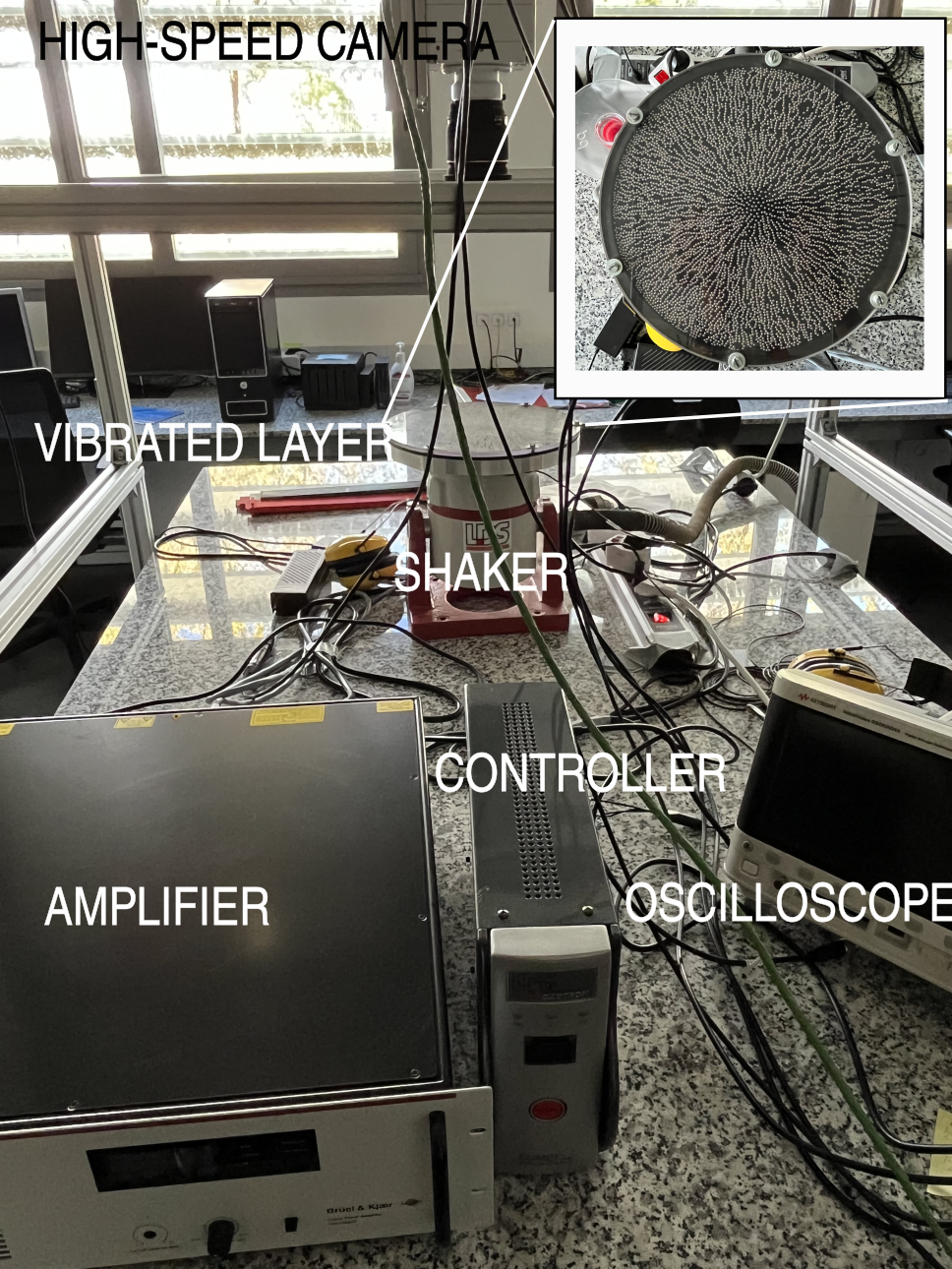}
  \caption{Experimental setup in our laboratory (Instituto de Computaci\'on Cient\'ifica Avanzada,
    ICCAEx, Badajoz, Spain). Inset: top-view close-up of the granular layer. \label{fig:exp_set-up_pic}}
\end{figure}

The granular material consists of stainless steel spheres (AISI~52100) disposed over a circular plate of
anodized aluminum. The particle-particle with coefficient of restitution is
$\alpha=0.95$ \cite{Louge_1999}. The gap spacing (system height) is $h=1.75\,\sigma$ is obtained with the use
of stainless steel ring-shaped spacers so that their inner diameter is $L = 256~\mathrm{mm}$. The
driving frequency is $\nu = 180~\mathrm{Hz}$. A photograph of the setup is shown in
Fig.~\ref{fig:exp_set-up_pic}. The spheres have a diameter
$\sigma =  3/32~\mathrm{in} \simeq 2.38252~\mathrm{mm} \pm 0.5~\mathrm{\mu m}$, mass density $\rho = 7833.4~\mathrm{kg/m^3}$, and mass
$m = 0.0555~\mathrm{g}$. The 2D particle density in the experiments is
$\rho_{2D} = 0.45$. The density definition follows previous studies \cite{Melby_2005,Reyes_2008}:
$\rho_{2D}\equiv N /N_\mathrm{max}$, where $N_\mathrm{max} = 10469$ is the maximum number of balls used in the
experiments that can fit in a hexagonally arranged 2D lattice in the circular plate of our set-up; i.e.,
$N_\mathrm{max} \sigma^2/L^2 \equiv \pi/(2 \sqrt{3})$ and thus, in the experiments, we use $N = 4711$ particles.

Movies of the experiments are recorded through the transparent lid. With suitable lighting, particle
horizontal positions are extracted and tracked with high accuracy \cite{Olafsen_1998}, so that the 2D
horizontal dynamics can be analyzed. We use the Crocker--Grier algorithm \cite{Crocker_1996}, in the Python
implementation of Allan \emph{et al.}~\cite{Allan_2025}, modified to suit our needs.

\textit{Protocol preparation.—} The protocol follows Appendix~A. Steps~1 and 2 simply require driving the
system at the desired acceleration, maintained by the closed-loop controller. In step~3, the layer is driven
at $\Gamma_0 = 7.0$ and then switched to $\Gamma_s = 3.15$ after a sufficiently long time. The entire process is
recorded, enabling extraction of the horizontal granular temperature. The control signals from the vibration
system and camera are fed to the oscilloscope, allowing identification of the frame in which the acceleration
change occurs. The camera resolution is 1\,ms (1000\,fps), and the vibration refresh time during this step is
$1/f = 5.5~\mathrm{ms}$, allowing accurate synchronization.

The intermediate steady temperatures $\{T_s^i\}$ from step~2 are used to identify the frames in step-3 at
which the corresponding temperatures are reached, which is needed for steps~4--5. Experimentally, these steps
proceed by preparing the system in the steady state at $\Gamma_0 = 7.0$, switching to $\Gamma_s = 3.15$, waiting
the required number of cycles, and then switching to the appropriate intermediate $\Gamma_w$. Reference times
for all switches are obtained from the oscilloscope traces, as in step~3.

\section{Appendix B. Description of the simulations}
\label{app:simulation}

The algorithm for molecular dynamics simulations has been used elsewhere \cite{Melby_2005}. The boundaries are
configured so that there is a pair of parallel vibrating walls that are perpendicular to gravity (Z direction)
$g=9.8~\mathrm{m/s^2}$. The system is periodic in the horizontal directions (XY coordinates).  The particle
interact  through a combination of a conservative restoring force and normal and tangential frictional forces,
in such a way that particle-particle forces can be expressed as

\begin{align}
  &\mathbf{F}_{ij}^\mathrm{rest}=Ym_i(|\mathbf{r}_{ij}|-\sigma)\mathbf{\hat{r}}_{ij}, 
  \mathbf{F}_{ij}^\mathrm{diss}=-\gamma_{n}m_i\mathbf{v}_{ij}^n, \nonumber \\
  &\mathbf{F}_{ij}^\mathrm{shear}=-\gamma_{s}m_i\mathbf{v}_{ij}^t. \nonumber \label{eq:particleforcesMD}
\end{align}

In the above equations, subscripts $i,j$ stand for particles; $m_i$ is the
particle mass; $\mathbf{r}_{ij}$ are relative positions: $\mathbf{r}_{ij}=\mathbf{r}_i-\mathbf{r}_j$, and
$\mathbf{\hat r}_{ij}=\mathbf{r}_{ij}/|\mathbf{r}_{ij}|$, with $F_{ij} = 0$  if $r_{ij} =|\mathbf{r}_{ij}| > \sigma$.  Analogously, $\mathbf{v}_{ij}$ stands for relative
velocities, and $\mathbf{v}_{ij}^n$, $\mathbf{v}_{ij}^t$ stand for the projections of relative velocities in
the normal and tangential directions respectively. The
coefficient $Y$ is the Young modulus that characterizes the (conservative) restoring force whereas $\gamma_n$,
$\gamma_s$ account for the dissipation in the normal and tangential directions respectively. There is also a
set of analogous particle-wall interactions, but with coefficients $Y_w, \gamma_{nw}, \gamma_{sw}$. Analogous equations would rule for the wall-particle
interactions, but with coefficients $Y_w, \gamma_{nw}, \gamma_{sw}$ respectively, which can take in the code different values to those for particle-particle interactions. Analogously, particle-wall interactions are null if $r_{iw} > \sigma/2$, where $r_{iw}$ is the distance between the i-th particle and the wall. In this
work we have used values of parameters that mimic the behavior of metallic balls with a coefficient of normal
restitution $\alpha=0.95$ for steel balls \cite{Reyes_2008}. Also, in order to simplify, we have used the same
values of force parameters for the wall-particle interactions. Thus, use here $Y=Y_w=10^7$~s$^{-2}$,
$\gamma_n=\gamma_{nw}=200$~s$^{-1}$, $\gamma_s=\gamma_{sw}=100$~s$^{-1}$. See \cite{Nie_2000} for further
reference. Other simulation
parameters are $\nu = 643.217~\mathrm{Hz}$, and $h=2.00~\sigma$, where
  $\sigma = 3/64~\mathrm{in} \simeq 1.190625~\mathrm{mm}$, and the 2D number density is $\rho_{2D}=0.433$. We have made $100$ replicas of each relaxation curve, which minimizes statistical uncertainty to nearly vanishing.

Molecular dymanics simulations were performed trying to mimmic the experimental conditions \cite{Melby_2005},
by adjusting the friction coefficients to the values that approximately would correspond to steel spheres
\cite{Sun_2006}.

\section{Appendix C: Implementation of the Kovacs Protocol in Experiments and Simulations}
\label{app:protocol}


This section describes the procedure used to implement the Kovacs protocol, which is identical in the
experiments and simulations. The protocol consists of the following five steps:

\begin{enumerate}[label=\arabic*. ]
  \item Determine the two steady states associated with the initial and final input accelerations—the upper and
    lower steady states illustrated in Fig.~\ref{fig:Kovacs_types}. These values correspond to
    $\Gamma_0 = 7.0$ and $\Gamma_s = 3.15$ respectively in the experiments and to
    $\Gamma_0 = 12.5$ and $\Gamma_s = 5.0$ respectively in the simulations.
  \item Measure the steady states for the intermediate acceleration values used in the experiments,
    $\Gamma_w = \{6.5, 6.0, 5.5, 5.0, 4.5, 4.0\}$ (and in
    simulations, $\Gamma_w = \{11.5, 10.5, 9.5, 8.5, 7.5, 6.5, 5.5\}$), and obtain the corresponding steady
  horizontal temperatures $\{T_s^{*}(\Gamma_w)\}$ in both experiments and simulations.
\item Record statistical realizations of the relaxation dynamics toward the lower-acceleration steady state
  ($\Gamma_s = 3.15$ for experiments, $\Gamma_s = 5.0$ in the simulations), starting from the steady state at
  the upper acceleration ($\Gamma_0 = 7.0$ for the experiments, $\Gamma_0 = 12.5$ in the simulations). Each
  relaxation curve results from an average over 5 independent realizations for experiments and over $N_r=100$
  independent realizations for simulations. Each replica is started from a freshly prepared steady
  state. Error bars are not shown for clarity; the spread across independent realizations is expected to be
  negligible compared to the symbol size. 
  
\item Construct relaxation curves by selecting points along the relaxation obtained in step~3 whose final
  temperatures match $T = \{T_s^{*}(\Gamma_w)\}$, using the respective upper steady states ($\Gamma_0 = 7.0$
  for experiments, $\Gamma_0 = 12.5$ for simulations) as initial conditions.
  \item Generate the Kovacs response curves by initiating the system from the final states identified in step~4, by
    applying the intermediate accelerations and recording the evolution until the steady values
    $\{T_s^{*}(\Gamma_w)\}$ are reached.
\end{enumerate}



\end{document}